\documentclass[apjl]{emulateapj}
\usepackage{epsfig}
\usepackage{apjfonts}

\slugcomment{To appear in ApJ, 20 Dec 2005, vol 635 issue} 

\begin{document}

\title{The First Laser Guide Star Adaptive Optics Observations of
the Galactic Center:  Sgr A*'s Infrared Color and the Extended Red 
Emission in its Vicinity}

\author{
A. M. Ghez\altaffilmark{1,2}, 
S. D. Hornstein\altaffilmark{1}, 
J.Lu\altaffilmark{1},
A. Bouchez\altaffilmark{3}, 
D. Le Mignant\altaffilmark{3}, 
M. A. van Dam\altaffilmark{3},
P. Wizinowich\altaffilmark{3}, 
K. Matthews\altaffilmark{4}, 
M. Morris\altaffilmark{1},
E. E. Becklin\altaffilmark{1},
R. D. Campbell\altaffilmark{3},
J. C. Y. Chin\altaffilmark{3},
S. K. Hartman\altaffilmark{3},
E. M. Johansson\altaffilmark{3},
R. E. Lafon\altaffilmark{3},
P. J. Stomski\altaffilmark{3},
D. M. Summers\altaffilmark{3}
}


\altaffiltext{1}{Division of Astronomy and Astrophysics, UCLA, Los Angeles, CA 90095-1547; ghez, seth, jlu, morris, becklin@astro.ucla.edu}
\altaffiltext{2}{Institute of Geophysics and Planetary Physics, UCLA, 
Los Angeles, CA 90095-1565}
\altaffiltext{3}{W. M. Keck Observatory, 65-1120 Mamalahoa Hwy, Kamuela, HI 9674
3; abouchez, davidl, peterw, rcampbell, jchin, mvandam, shartman, erikj, 
rlafon, pstomski, dsummers@keck.hawaii.edu}
\altaffiltext{4}{Caltech Optical Observatories, California Institute of 
Technology, MS 320-47, Pasadena, CA 91125; kym@caltech.edu}

\begin{abstract}
We present the first Laser Guide Star Adaptive Optics
(LGS-AO) observations of the Galactic center.
LGS-AO has dramatically improved the quality and robustness with
which high angular resolution infrared images of the Galactic center
can be obtained with the W. M. Keck II 10-meter telescope.
Specifically, Strehl ratios of  
0.7 and 0.3 at L'[3.8 $\mu m$] and K'[2.1 $\mu$m], respectively,
are achieved in these LGS-AO images; these 
are at least a factor of two higher
and a factor of four to five more stable against
atmospheric fluctuations than the Strehl ratios delivered
thus far with the Keck
Natural Guide Star AO system on the Galactic center.
Furthermore, these observations
are the first that cover a large area ($76\arcsec \times 76\arcsec$)
surrounding the central black hole at diffraction-limited resolution
for an 8-10 meter class telescope.
During our observations, the infrared counterpart to the 
central supermassive black hole, Sgr A*-IR, showed significant infrared 
intensity variations,
with observed L' magnitudes ranging from 12.6 to 14.5 mag and
a decrease in flux density of a factor of two over an 8 minute interval.
The faintest end of our L' detections, 1.3 mJy (dereddened), is the lowest 
level of emission yet observed for this source by a factor of 3.
No significant variation in the location of SgrA*-IR is detected
as a function of either wavelength or intensity.  Previous claims
of such positional variations are easily attributable to
a nearby (0\farcs 09 or 720 AU, projected), 
extended, very red source, which we suggest arises 
from a locally heated dust feature.
Near a peak in its intensity, we obtained the first measurement of 
SgrA*-IR's K'-L' color;  its K'-L' of 
3.0 $\pm$ 0.2 mag (observed) or 1.4 $\pm$ 0.2 (dereddened) 
corresponds to an intrinsic spectral index of $\alpha$ -0.5 $\pm$ 0.3 for 
F$_{\nu} \sim \nu^{\alpha}$.  This is significantly bluer than other recent 
infrared measurements from the literature, which suggest $\alpha$ = -4 $\pm$ 1.  
Because our measurement was taken 
at a time when Sgr A* was $\sim$6 times brighter in the infrared than
the other measurements, we posit that the spectral index of 
the emission arising from the vicinity of our Galaxy's 
central black hole may depend on the strength of the flare, with stronger 
flares giving rise to a higher fraction of high energy electrons in the 
emitting region.

\end{abstract}


\section{INTRODUCTION}

High angular resolution infrared observations have had a large impact
on our understanding of the center of the Milky Way Galaxy. 
Application of speckle imaging techniques generated the first 
high angular resolution (0\farcs 05 - 0\farcs 1) 
images and demonstrated the existence of a 
supermassive ($\sim3.7 \times 10^6 M_{\odot}$) black hole
through measurements of orbital motion in the plane of the sky for 
stars within the central few arcseconds of our Galaxy 
(e.g., Eckart \& Genzel 1997; Ghez et al. 1998, 2000,
2003, 2005; Genzel et al. 2000; Eckart et al. 2002; 
Sch\"odel et al. 2002, 2003).
While these speckle imaging experiments resulted
in the most convincing case of a supermassive black hole
at the center of any normal type galaxy, the application
of this technique to the Galactic center has been limited
primarily to proper motion measurements at a single wavelength.
The advent of Adaptive Optics (AO) systems that are based on natural guide stars
(NGS) allowed spectroscopy with sufficient angular and spectral resolution
to detect weak absorption lines in individual stars orbiting in close 
proximity to the central black hole.  These observations revealed that
luminous stars located no more than 2000 AU from the central black hole
are OB stars, which are thought to be less than 10 Myrs old 
and which therefore raise the interesting question of how 
apparently young stars form in a region that is inhospitable to star 
formation (Ghez et al. 2003).  These observations, in concert
with the proper motion measurements, also provided a
direct measurement of the distance to the Galactic center 
(Eisenhauer et al. 2003).  NGS-AO systems also increased the
efficiency and depth of high angular resolution observations; this
led to the discovery of an infrared counterpart to
the central black hole, Sgr A*-IR 
(Genzel et al. 2003a; Ghez et al. 2004) and permitted 
detailed studies of the stellar number density distribution,
which suggested the existence of central stellar cusp (Genzel et al. 
2003b).  While NGS-AO has greatly enhanced the versatility of 
high angular resolution observations of the Galactic center,
the quality of NGS-AO observations degrades rapidly both with distance 
from the NGS and with dimness of the NGS. 

In this paper, we present the first Laser Guide Star Adaptive Optics
(LGS-AO) observations of the Galactic center.  LGS-AO has dramatically
improved the quality and robustness with which 
high angular resolution images can be obtained with the
W. M. Keck II 10-meter telescope, revealing new features in the
infrared emission coincident with and surrounding the central black hole,
clear short timescale variations in the infrared emission arising
from just outside the black hole, and the first measurement of SgrA*-IR's
K'-L' color.
Furthermore, these observations are the first that can cover a large area 
surrounding the central black hole at diffraction-limited resolution. 
These measurements represent an exciting next step in our understanding
of the central black hole and its surrounding as well as in our
technical ability to obtain high angular resolution data.

\section{OBSERVATIONS}

Images of the Galactic center
were obtained on 2004 July 26 (UT) with the W. M. Keck II 10-meter
telescope using the new facility Laser Guide Star Adaptive Optics 
system (Wizinowich et al., in prep; van Dam et al., in prep) and the 
near-infrared
camera, NIRC2 (Matthews, in prep).  During these observations,
an artificial guide star was 
generated with a pulsed dye laser, which was tuned to the Na D atomic
transition (589 nm) and which was run at 14 W output power.
At the beginning of the night, the size of the LGS, imaged at zenith with the
whole telescope pupil, was measured to be 1".3 x 2".1, elongated due to the
finite thickness of the sodium layer. The size of the LGS across the
wavefront sensor varied from 1".2 x 1".4 close to the location of the LGS
launch telescope to 1".5 x 3".1 for the spots furthest from launch
telescope. Although the intrinsic LGS spot size increases with increasing
airmass due to additional turbulence on its upward and downward trajectory,
the spot elongation decreases, since the LGS at a larger distance from the
telescope.  Over the course of our observations,
the LGS brightness was equivalent to 
an V magnitude of 11.4 star and it varied by only 5\% (1$\sigma$).
For all the observations reported here,
the LGS's position was always locked to the center of the 
NIRC2 field of view.  
While observations of the artificial guide star provide the necessary
information to correct most of the important atmospheric aberrations, 
it does not provide information on the tip-tilt term, which, for these observations, 
was obtained from visible observations of USNO 0600-28577051 
(R = 13.7 mag and $\Delta r_{SgrA*}$ = 19$\arcsec$).  
The science instrument houses
a 1024$\times$1024 InSb array, which has pixel scales of
9.93 and  39.9 milli-arcsec per pixel 
and corresponding field of view of
10\farcs 2$\times$10\farcs 2 and 40\farcs 8$\times$40\farcs 8, 
respectively.  All observations were made using correlated double sampling readout
of the array, which produced an RMS readout noise of 77 electrons.
With the LGS-AO/NIRC2 setup,
two sets of eight 30 sec narrow-field images, each composed of 120 coadded 
0.25 sec 
exposures, were obtained through the 
L'($\lambda_o$=3.78 $\mu$m, $\Delta \lambda$=0.70 $\mu$m) photometric bandpass;
each set of eight images was composed of four pairs of images centered
at the corners of a 0\farcs 5 $\times$ 0\farcs 5 box.
An additional twelve 9 sec 
K'($\lambda$=2.12 $\mu$m, $\Delta \lambda$=0.35 $\mu$m) 
narrow-field images, each composed of 50 coadded 0.181 sec exposures, were 
collected.
In the wide field-of-view mode, 
two sets of 
five 25 sec (100 coadded 0.25 exposures) narrow bandpass filter
(set 1 [He-I filter]: $\lambda_o$ = 2.06 $\mu$m, $\Delta \lambda$ = 0.03 $\mu$m
and
set 2 [narrow band continuum filter]: $\lambda_o$ = 2.27 $\mu$m, $\Delta \lambda$ = 0.03 $\mu$m)
images, dithered by 18$\arcsec$ to create a $76\arcsec \times 76\arcsec$ 
field-of-view, were also obtained. 
A similar sequence on a relatively dark portion of the sky was
observed both before and after the Galactic center observations
to measure the background emission.

The operational efficiency of the LGS-AO system 
was quite high.  Before these observations could begin,
roughly 5 min were required to acquire the tip-tilt star and 
propagate the laser.  After this initial set-up time, 
the only other overhead introduced by the LGS-AO system
was an additional $\sim$20 sec with each dither to move the 
laser to the new field center.  Wizinowich et al. (2005, in prep)
provide further details on the science observing efficiency for 
the LGS-AO system in all its possible observing modes.

During these measurements, the uncorrected 2 $\mu$m seeing disk size 
for observations of the Galactic center was $\sim$0\farcs 5, which is 
close to the median value of 0\farcs 4 for the Keck telescopes.
On-axis K' LGS-AO observations of the tip-tilt star, which were taken just 
before the observations centered on Sgr A* at an airmass (AM) of 1.53, 
yielded a point spread function
with a Strehl ratio of 0.422 and a FWHM of 52.3 mas.  Compared
to the LGS-AO performance for observations centered on SgrA*, which
yield an average K' Strehl ratio of 0.31 (see \S4 for detailed
description of LGS-AO performance), this implies an atmospheric isokinetic
angle\footnote{The isokinetic angle, $\theta_{TA}$, is the 
characteristic angular offset at which the RMS wavefront error
due to tilt decorrelation ($\sigma_{TA}$) is 1 radian 
(${\sigma_{TA}}^2 = (\theta / \theta_{TA})^2$, where $\theta$ is the
distance from the tip-tilt star; Hardy 1998).} 
of $34\arcsec (1.53/AM)^{1.5}$ for these observations and
64\arcsec for observations at zenith.  

\begin{deluxetable*}{llccccc}
\tablewidth{0pt}
\tablecaption{Summary of Key Detections in the Average LGS-AO Maps from 2004 July 26 (UT)\label{tab_srcs}}
\tablehead{
	\colhead{Name} &
	\colhead{Other} &
	\colhead{r} &
	\colhead{$\Delta \alpha_{dyn. ~ cent.}$} &
	\colhead{$\Delta \delta_{dyn. ~ cent.}$} &
	\colhead{K'} &
	\colhead{L'}
\\
	\colhead{} &
	\colhead{Name\tablenotemark{a}} &
	\colhead{(arcsec)} &
	\colhead{(arcsec)} &
	\colhead{(arcsec)} &
	\colhead{(mag)} &
	\colhead{(mag)}
}
\startdata
SgrA*-IR & \nodata & 0.00 &  $~$0.00 &  $~$0.00 & 15.99 $\pm$ 0.06 & 13.39 $\pm$ 0.09 \\
Extended Red & \nodata & 0.10 & -0.07 & -0.07 & $<$16.2	   & 13.64 $\pm$ 0.15 \\
Emission     &	       &   & 	      &	      &	 	 &	\\
S0-17	& \nodata  & 0.10 &  $~$0.03 & -0.10 & 15.52 $\pm$ 0.06 & 13.93 $\pm$ 0.12 \\
S0-2	& S2 	   & 0.12 &  $~$0.03 &  $~$0.12 & 14.17 $\pm$ 0.03 & 12.78 $\pm$ 0.03 \\
S0-20   & S13 	   & 0.14 & -0.11 &  $~$0.08 & 15.74 $\pm$ 0.20 & 14.18 $\pm$ 0.07 \\
S0-19 &	 S12	   & 0.32 & -0.06 &  $~$0.31 &  15.40 $\pm$ 0.08 & 13.92 $\pm$ 0.03 \\
S0-8 &	ID14	   & 0.37 & -0.28 &  $~$0.24 &  15.68 $\pm$ 0.07 & 14.01 $\pm$ 0.05 \\
\enddata
\tablenotetext{a}{Other names from Sch\"odel et al.\ (2003).}
\end{deluxetable*}

\section{DATA ANALYSIS}

The standard image reduction steps of background subtraction, flat fielding,
bad pixel repair, and optical distortion correction were carried out on each 
image.  Since the background in the L' images is dominated by the
thermal emission from the AO system, the best background subtraction 
was achieved by subtracting from each image the average of 
sky images that were taken with as similar a rotator angle
as possible.  
The optical distortion corrections applied were taken from
the NIRC2 pre-ship review results 
(http://alamoana.keck.hawaii.edu/inst/nirc2/) and are relatively small
for the narrow field of view camera.  The pre-ship review results do 
not adequately describe the optical distortions in the wide field
of view camera.  We therefore restrict our analysis of the wide field
data set to the small region around SgrA*-IR and will report the
results of our search for stars with He-I line features, which 
are more sensitive to these field variations, elsewhere (Lu et al., 
in prep.)

Point sources were identified and characterized in both 
the individual images and in an average of all the images
by running them through the point spread function (PSF) fitting
program, StarFinder (Diolaiti et al. 2000).   This
generated an estimate of each image's PSF based on the
several bright point sources (IRS 16C, 16NW, 29N, 16NE, and 33E for the K' 
images and IRS 16C, 16NW, and 29NE in the L' images)
and relative flux densities for all sources that 
were identified.  Using the apparent magnitudes measured
by Blum et al. 1996 at K and Wright et al. (in prep) at L'
for IRS 16NE (K=9.00, L'=7.37 mag),
IRS 16NW (K=10.03, L'=8.43 mag), and IRS 16C (K=9.83, L'=8.14 mag),
we derived apparent magnitudes for all the sources
detected with StarFinder and estimated the uncertainties in the calibration
to be $\sim$4\% and 5\% for the K' and L' observations, respectively\footnote
{The formal uncertainty for the K' absolute magnitudes is only 2\%,
but we increase it to 4\% to accommodate the differences between the
K and K' filters.  Nonetheless, our final reported uncertainties are
dominated by the relative flux measurements as opposed to this
absolute calibration uncertainty.}.  
The apparent magnitudes were de-reddened 
assuming a visual extinction of 29 $\pm$ 1 mag from Wright et al. 
(in prep) and an extinction law derived by Moneti et al. (2001),
and then converted to flux densities with zero points 
from Tokunaga 2000 (see Table~\ref{tab_sum}'s footnotes for values).

Extra steps were taken with StarFinder to ensure that SgrA*-IR, which
is located in a very crowded region, was accurately measured in 
the individual short exposure frames. 
First, StarFinder was run using a detection threshold of 0.6 for the 
correlation between a source and the PSF.  Second, StarFinder was run
in a mode in which it fits only sources from an input list.  Our list
included all sources detected from the first step along with 
any source within 0\farcs 15 of the dynamical center that
was detected in the average map at that wavelength, 
but was missed in the first step for that individual short exposure
\footnote{The following sources within the central 0\farcs 15 were missed 
in the first step: 
SgrA*-IR in 5 K' and 8 L' maps,
S0-17 in 1 K' and 4 L' maps, and S0-20 in 1 K' and 4 L' maps.}.
Since
StarFinder's PSF fitting routine is very sensitive to local
confusion, these added sources, if overlooked, can affect the 
resulting photometry.
Since Frame 82 had an anomalously poor-quality PSF and StarFinder failed to 
identify any of the sources within 0\farcs 1 of SgrA*-IR in this first step, it was 
dropped from any further analysis.
The second step recovered all the missing detections of these central 
sources except for SgrA*-IR in 3 K' maps and S0-20 in one L' map.
The consecutive images \#113 and 114 with no detections of SgrA*-IR 
were averaged together
and analyzed as described above, but still led to a non-detection of SgrA*-IR.

Uncertainties in SgrA*-IR's photometry 
were estimated from the flux density fluctuations between the maps 
for stars that are detected in at least 75\% of the maps.  Since
the stars are not variable on the timescale of this experiment 
and the image quality is fairly constant
across the maps, the rms of these stars' flux densities provide an 
estimate of the
measurement uncertainty as a function of flux density. 
We therefore set the uncertainty in a given SgrA*-IR flux density 
measurement equal to the average rms of stars whose flux density differs by 
no more than 1 mJy from that measurement of SgrA*-IR.
Between 
10 and 75 stars contribute to most of these uncertainty estimates, depending
on the brightness of SgrA*-IR.

For the two K' maps without detections of SgrA*-IR
(\#83 and the average of \#113 and 114), upper limits on SgrA*-IR's
flux density 
are derived from aperture photometry. 
In order to remove the effects of nearby stars in such a crowded region,
we subtracted, from the original map, the model created by StarFinder 
of all the detected point sources, 
with the exception of  
two sources at r$>$0\farcs 1 that served as photometric reference 
sources (S0-19 \& S0-8; see Table~\ref{tab_srcs}).  
Aperture photometry was then conducted on this difference map,
using an aperture radius of $\sim$0\farcs 06
and a surrounding sky annulus of width $\sim$0\farcs 03,
at the locations of SgrA*-IR, the reference sources, and a few background
locations near SgrA*-IR.  These were combined to generate 3$\sigma$
detection limits.  A similar analysis was carried out on the two wide-field
maps that contained SgrA*-IR.

\begin{figure*}
\epsscale{1.0}
\plotone{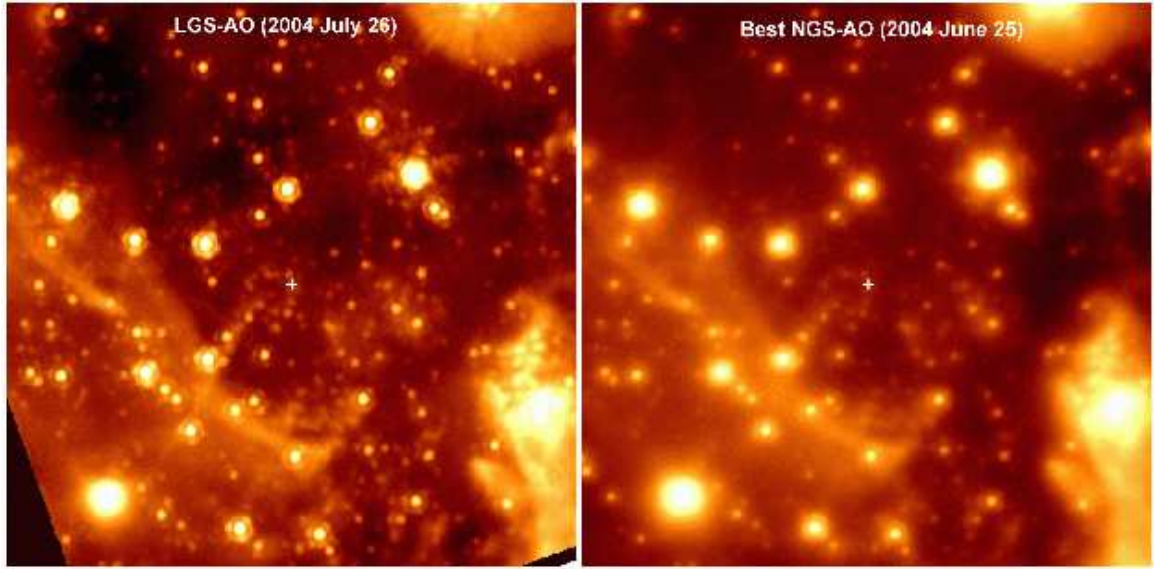}
\figcaption{
A comparison of the first LGS-AO image (left) and the best NGS-AO image
(right)
taken with the W. M. Keck II 10 m telescope (2002-2004)
in the L'(3.8 $\mu$m) photometric bandpass of the 
central 7\farcs5$\times$7\farcs5 of our Galaxy.  In both images, the cross denotes the location 
of the central supermassive black hole and the orientation is North up
and East to the left.  The LGS-AO image  
has a Strehl ratio that is a factor of two higher than 
that obtained in the NGS-AO image; furthermore, the LGS-AO image resulted from
an exposure time of only 8 min, a factor of $\sim$ 20 less than the
comparison NGS-AO image.  The LGS-AO system has therefore
dramatically improved the image quality that can be obtained
on the Galactic center with the Keck telescope.
\label{lgs_img}
}
\end{figure*}

\begin{figure*}
\epsscale{1.0}
\plotone{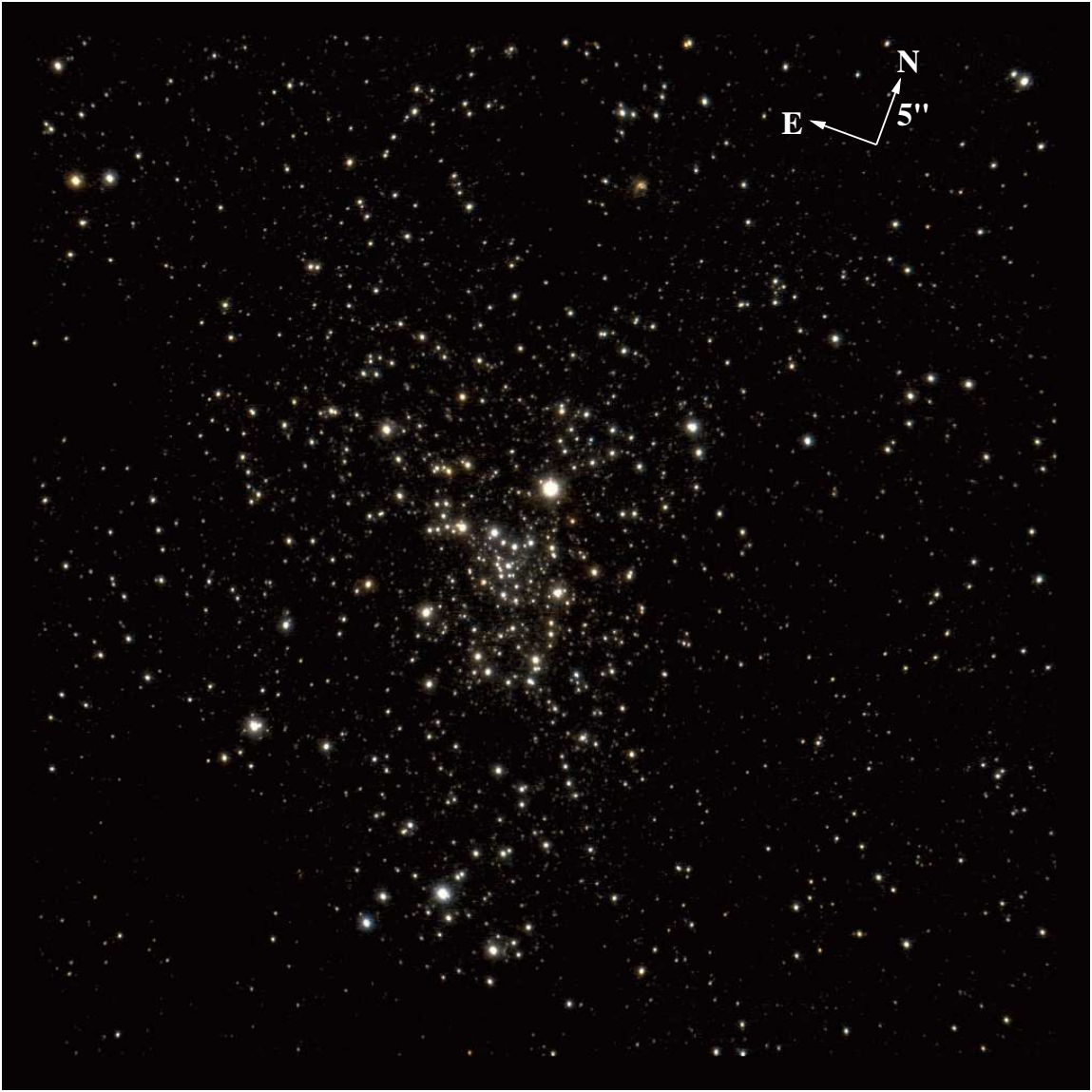}
\figcaption{
A $76\arcsec \times 76\arcsec$ diffraction-limited 2 $\mu$m 
mosaic of the Galactic center obtained with five pointings
of the LGS-AO system.  This is the widest near-infrared field
imaged yet at this angular resolution.
}
\end{figure*}

\begin{deluxetable*}{cccccccccccc}
\tabletypesize{\scriptsize}
\tablewidth{0pt}
\tablecaption{Summary of 2004 July 26 (UT) LGS-AO Time-Resolved Observations of
SgrA*-IR\label{tab_sum}}
\tablehead{
        \colhead{File \#}&
        \colhead{Time}&
        \colhead{T$_{exp}$}&
        \colhead{$\sigma_{wf}$\tablenotemark{a}}&
        \multicolumn{2}{c}{FWHM (mas)} &
        \multicolumn{2}{c}{Strehl Ratio } &
        \multicolumn{2}{c}{K' Brightness}   &
        \multicolumn{2}{c}{L' Brightness}  
\\
	\colhead{} & 
	\colhead{(UTC)} & 
	\colhead{(sec)} & 
	\colhead{(nm)} & 
	\colhead{K'} & 
	\colhead{L'} & 
	\colhead{K'} & 
	\colhead{L'} & 
        \colhead{Observed Mag.}&
        \colhead{F$_{\nu}$ (mJy)\tablenotemark{b,c}}&
        \colhead{Observed Mag.}&
        \colhead{F$_{\nu}$ (mJy)\tablenotemark{b,c}}
}
\startdata
77   & 08:18:50 & 9  & 344 & 66 & \nodata & 0.36 & \nodata & 15.85 $\pm$ 0.22 &  5.9 $\pm$ 1.2 & \nodata & \nodata  \\
78   & 08:19:58 & 9  & 340 & 63 & \nodata & 0.36 & \nodata & 16.10 $\pm$ 0.32 &  4.7 $\pm$ 1.4  & \nodata & \nodata \\
79   & 08:21:34 & 9  & 346 & 62 & \nodata & 0.35 & \nodata & 15.50 $\pm$ 0.16 &  8.1 $\pm$ 1.2 & \nodata & \nodata \\
80   & 08:21:59 & 9  & 343 & 62 & \nodata & 0.36 & \nodata & 15.64 $\pm$ 0.18 &  7.1 $\pm$ 1.2 & \nodata & \nodata \\
81   & 08:22:53 & 9  & 427 & 83 & \nodata & 0.20 & \nodata & 15.23 $\pm$ 0.15 & 10.4 $\pm$ 1.4 & \nodata & \nodata \\
83   & 08:24:04 & 9  & 387 & 73 & \nodata & 0.27 & \nodata & $>$15.08\tablenotemark{c}   & $<$12.0  & \nodata & \nodata\\
84   & 08:24:30 & 9  & 346 & 62 & \nodata & 0.35 & \nodata & 15.57 $\pm$ 0.17 &  7.6 $\pm$ 1.2 & \nodata & \nodata \\
85   & 08:25:14 & 9  & 382 & 72 & \nodata & 0.29 & \nodata & 15.77 $\pm$ 0.20 &  6.3 $\pm$ 1.2 & \nodata & \nodata \\
86   & 08:25:40 & 9  & 328 & 59 & \nodata & 0.39 & \nodata & 15.63 $\pm$ 0.18 &  7.2 $\pm$ 1.2 & \nodata & \nodata \\
87   & 08:27:04 & 30 & 378 & \nodata & 81.1 & \nodata & 0.67 & \nodata & \nodata & 12.58 $\pm$ 0.09 &  9.74 $\pm$ 0.79 \\ 
88   & 08:28:02 & 30 & 378 & \nodata & 81.9 & \nodata & 0.67 & \nodata & \nodata & 12.64 $\pm$ 0.08 &  9.21 $\pm$ 0.68 \\ 
89   & 08:29:31 & 30 & 329 & \nodata & 80.6 & \nodata & 0.74 & \nodata & \nodata & 12.81 $\pm$ 0.10 &  7.88 $\pm$ 0.71 \\ 
90   & 08:30:29 & 30 & 316 & \nodata & 80.3 & \nodata & 0.76 & \nodata & \nodata & 12.90 $\pm$ 0.11 &  7.25 $\pm$ 0.76 \\ 
91   & 08:31:46 & 30 & 337 & \nodata & 80.4 & \nodata & 0.73 & \nodata & \nodata & 12.89 $\pm$ 0.12 &  7.32 $\pm$ 0.78 \\ 
92   & 08:32:45 & 30 & 347 & \nodata & 80.6 & \nodata & 0.72 & \nodata & \nodata & 12.84 $\pm$ 0.10 &  7.66 $\pm$ 0.72 \\ 
93   & 08:34:02 & 30 & 360 & \nodata & 80.8 & \nodata & 0.70 & \nodata & \nodata & 12.99 $\pm$ 0.12 &  6.67 $\pm$ 0.74 \\ 
94   & 08:35:00 & 30 & 370 & \nodata & 81.8 & \nodata & 0.68 & \nodata & \nodata & 13.19 $\pm$ 0.14 &  5.55 $\pm$ 0.71 \\ 
95   & 08:39:19 & 25 & \nodata & \nodata & \nodata & \nodata & \nodata & $>$16.66\tablenotemark{c,d} & $<$2.8 & \nodata & \nodata \\ 
104   & 08:49:42 & 25 & \nodata & \nodata & \nodata & \nodata & \nodata & $>$16.31\tablenotemark{c,d} & $<$3.5 & \nodata & \nodata \\ 
113-4 & 09:00:19 & 18 & 371 & 65.2 & \nodata & 0.30 & \nodata & $>$16.00\tablenotemark{c} & $<$5.1 & \nodata & \nodata \\
115  & 09:02:10 & 30 & 404 & \nodata & 82.7 & \nodata & 0.64 & \nodata & \nodata & 14.31 $\pm$ 0.19 & 1.98 $\pm$ 0.35  \\
116  & 09:03:08 & 30 & 403 & \nodata & 81.8 & \nodata & 0.64 & \nodata & \nodata & 14.54 $\pm$ 0.21 & 1.60 $\pm$ 0.30 \\
117  & 09:04:22 & 30 & 399 & \nodata & 81.6 & \nodata & 0.64 & \nodata & \nodata & 14.52 $\pm$ 0.20 & 1.63 $\pm$ 0.30 \\
118  & 09:05:28 & 30 & 405 & \nodata & 82.0 & \nodata & 0.64 & \nodata & \nodata & 14.56 $\pm$ 0.21 & 1.57 $\pm$ 0.30 \\
119  & 09:06:45 & 30 & 394 & \nodata & 80.4 & \nodata & 0.65 & \nodata & \nodata & 14.56 $\pm$ 0.21 & 1.57 $\pm$ 0.30 \\
120  & 09:07:44 & 30 & 419 & \nodata & 82.0 & \nodata & 0.62 & \nodata & \nodata & 14.78 $\pm$ 0.26 & 1.28 $\pm$ 0.30  \\
121  & 09:09:04 & 30 & 372 & \nodata & 81.3 & \nodata & 0.68 & \nodata & \nodata & 14.74 $\pm$ 0.25 & 1.33 $\pm$ 0.30 \\
122  & 09:10:02 & 30 & 377 & \nodata & 82.2 & \nodata & 0.67 & \nodata & \nodata & 14.73 $\pm$ 0.25 & 1.34 $\pm$ 0.30 \\
\enddata
\tablenotetext{a}{$\sigma_{wf}$ is the RMS wavefront error}
\tablenotetext{b}{The values given for flux densities are de-reddened flux
densities and derived from the apparent magnitudes with the 
following assumptions: (1) a visual extinction of 29$\pm$1 mag from 
Wright et al. (in prep) and an extinction law derived by Moneti et al. (2001; 
$\frac{A_{2.2 \mu m}}{A_V}$ = 0.1108;
$\frac{A_{3.8 \mu m}}{A_V}$ = 0.0540),
which gives a K' extinction of 3.21 mag and a L' extinction of 1.57 mag, and
(2) zero points (i.e., the flux density of a 0 mag star) 
for the K' and L' magnitude scales of 667 and 248 Jy,
respectively (Tokunaga 2000).}
\tablenotetext{c}{The uncertainties in the de-reddened flux densities do
not explicitly include the de-reddening uncertainties, which are $\sim$5\%
at K' and $\sim$11\% at L'.  For our measurements, these are only 
significant at L'.}
\tablenotetext{c}{Three-$\sigma$ limits}
\tablenotetext{d}{These values are from the data sets obtained with the
wide-field camera (and hence we do not attempt to report the PSF 
characterstics) and narrow band-pass filters.}
\end{deluxetable*}

\begin{figure}
\epsscale{1.0}
\plotone{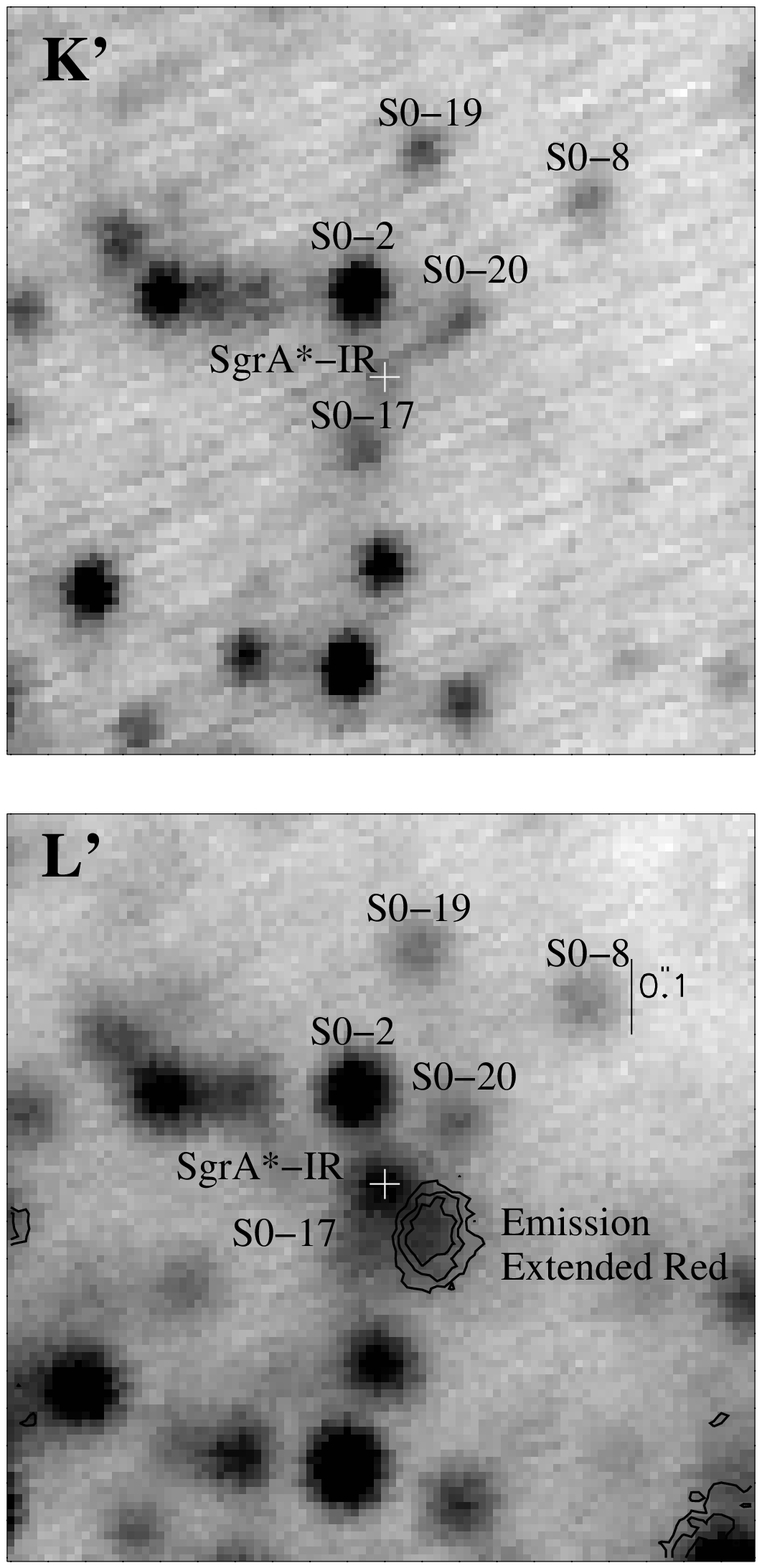}
\figcaption{
The central $1\arcsec \times 1\arcsec$ region of the average of all the
K' (left) and L' (right) LGS-AO images.  
The position of Sgr A*-IR is marked with a white cross and 
the labeled sources are those that are either within 0\farcs 15 of the 
dynamical center (Extended Red Emission, S0-17, S0-2, and S0-20)
or that were used to establish upper limits for detection (S0-19 and S0-8).
The contour image is what remains after all subtracting out all identified 
point sources, leaving primarily just the emission from
the extended source.  North is up and East is to the left. 
\label{extended}
}
\end{figure}

\section{LGS-AO PERFORMANCE}

The laser guide star adaptive optics system has 
dramatically improved the versatility, image quality, 
and robustness of adaptive optics observations of the 
Galactic center.  Based on the PSFs estimated from StarFinder (see \S3), 
the images all have consistently high 
Strehl ratios, 0.68 $\pm$ 0.04 for the L'(3.8 $\mu$m)-band 
images and 0.31 $\pm$ 0.06 for the K'(2.1 $\mu$m)-band images,
as well as PSF full width at half maxima (FWHM) that are also
all consistently very nearly diffraction-limited;
at L' the FWHM is 80.6 $\pm$ 0.7 mas 
and at K' the FWHM is 63 $\pm$ 8 mas.
Using the extended Mar\'echal approximation that $S = e^{-(2 \pi \sigma_{wf} / \lambda )^2}$,
where $S$ is the Strehl ratio, $\sigma_{wf}$ is the RMS wavefront error
and $\lambda$ is the observing wavelength, we find an average
RMS wavefront error of 371 $\pm$ 30 nm for these observations. 
Compared to the Keck natural guide star adaptive optics system (NGS-AO), 
the LGS-AO system can observe the Galactic center under 
much poorer atmospheric conditions and delivers L' images that have,
on average, a factor of 2 higher Strehl ratios 
and are a factor of 4-6 more stable; specifically, 
during our best L' NGS-AO nights (2002-2004)
images have Strehl ratios of 0.33 $\pm$ 0.06 and PSF FWHM of 
93 $\pm$ 5, or equivalently $\sigma_{wf}$ of 620 $\pm$ 50 nm  
(see Ghez et al. 2004 and Hornstein et al., in prep).  
This improvement is depicted in Figure 1, which compares the average of 
all our L' LGS-AO images to the best NGS-AO L' image that we 
obtained at the Keck telescope between 2002 and 2004; the improvement 
stems primarily from
having a much brighter, by about 2 mag, on-axis source to correct for the 
non-tip/tilt 
atmospheric aberrations with the LGS-AO system, compared to the faint
(R=13.2 mag) off-axis (30$\arcsec$) star used for the NGS-AO observations.  
The LGS-AO image is the highest quality L' image 
obtained thus far of the center of our Galaxy.  
Somewhat better K' performance
has been obtained with the VLT's NGS AO system since it has an infrared
wavefront sensor to take advantage of the bright, very nearby, 
infrared source IRS7 (K=6.4
mag) and the Galactic center passes directly overhead (versus a 50 deg
zenith angle from Hawaii).  However, this is only possible within the
isoplanatic angle ($\sim$20$\arcsec$ at 2$\mu$m) of this source, since no 
other comparably bright infrared source exists in this region.  
Figure 2 shows the remarkably large field coverage that is now possible
with the LGS-AO system.  This $76\arcsec \times 76\arcsec$ 
region is the largest area imaged at the diffraction limit of a 8-10 meter 
telescope by a factor of 4 and has a much more uniform point spread
function over comparable areas.  This approach is limited only by 
the isokinetic angle
($\sim$40-75$\arcsec$ at 2 $\mu$m), which is much larger than the isoplanatic angle, and by the
availability of tip-tilt stars, which need not be as bright as the
natural guide stars necessary to correct all the aberrations, and
which are therefore much more plentiful.

\section{RESULTS}

The improved image quality greatly facilitates detailed study of 
the infrared emission arising in the crowded region in the vicinity 
of our Galaxy's central black hole.
Within a radius of 0\farcs 15 of the dynamical center of our Galaxy, 
there are three known proper motion sources (S0-2, S0-17, S0-20; 
Ghez et al. 2005),
all of which are detected in both the K' and L' LGS-AO images
(see Table~\ref{tab_srcs} and Figure~\ref{extended}).  
In addition, the L' LGS-AO image reveals a new source (L' $\sim$ 13.3 mag), 
whose photocenter is located 0\farcs 094 South-West from Sgr A*-IR
(see Figure~\ref{extended} (left)).
This source, which was originally reported by Ghez et al. (2004) and 
has since been independently reported by Cl\'enet et al. (2005; labeled
SgrA*-f), can be 
equally well modeled as two roughly equal point
sources separated by 0\farcs 057 or as a two-dimensional gaussian that 
has a major-axis of 120 mas (corresponding to an intrinsic size of $\sim$
80 mas). No source is detected at this location
in the K' LGS-AO image down to $\sim$ 16.1 mag (2$\sigma$), which implies that
its (K'-L') must be redder than 2.8 mag (observed) or 1.2 mag (de-reddened).
Combining the photocenter of this LGS-AO detection with the photocenter
detections of this source in our earlier NGS-AO maps (Ghez et al. 2004
[2002 May 31 \& 2003 June 10, 16-17];
Hornstein et al., in prep [2004 June 25-28 \& 2004 July 6-7]), which do not
spatially resolve the source, we find its proper motion to be statistically 
consistent with zero to within 2$\sigma$ (9.6 $\pm$ 4.1 mas yr$^{-1}$).

\begin{figure}
\epsscale{1.0}
\plotone{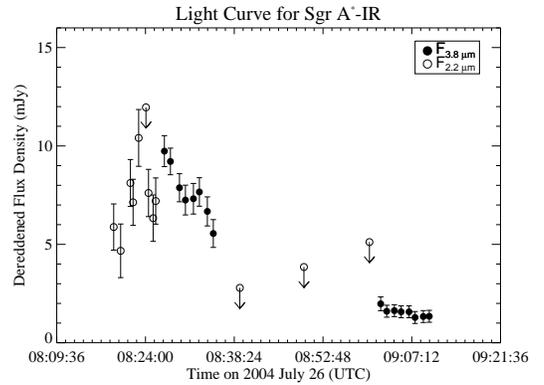}
\figcaption{
Light curve for Sgr A*-IR.  
The near simultaneity (within
1 minute) of the K' and L' measurements at $\sim$8:26 (UT) provide the 
first measurement of Sgr A*-IR's K'-L' color.
Significant flux density variations are also detected
both within the K' (unfilled points) and L' (filled points) measurements.
The emission from Sgr A*-IR is observed to increase, peak, and then
decrease by a factor of two in 8 minutes.  Another $\sim$30 minutes later 
it is observed to have a flux density of only 1.3 mJy at 3.8 $\mu$m.
This is the lowest level Sgr A*-IR has ever been detected by
a factor of 3, suggesting that no true steady-state level has yet 
been observed.
\label{lightcurve}
}
\end{figure}

In spite of the stellar crowding, Sgr A*-IR is also clearly detected in 
both the average K' and L' images as well as in the majority of 
the individual L' and K' images (see Table~\ref{tab_sum} 
and Figure 3).   
At the beginning of our 
observations, Sgr A*-IR at K' appears to brighten and reach a peak of
K'$_{obs} \sim$15.5. 
At the end of the K' sequence, Sgr A* has a K'$_{obs}$ 
magnitude of 15.6 $\pm$ 0.2 or, equivalently, a de-reddened flux density
of 7 $\pm$ 1 mJy.  After a time gap of $\sim$1 min, its 
observed L' magnitude is 12.58 $\pm$ 0.09 (F$_{3.8 \mu m, de-reddened}$ = 
9.7 $\pm$ 0.8 mJy).  This provides an estimate of the color of
Sgr A* of (K'-L')$_{obs}$ = 3.0 $\pm$ 0.2 or (K'-L')$_{de-reddened}$ = 1.4 $\pm$ 0.2. 
Over the course of only 8 minutes, its flux density decreases by a factor of 2
and, another 27 minutes later,
Sgr A* is a factor of 3 fainter still, or 1.3 mJy (de-reddened), in the 
L' photometric bandpass.  This is the faintest Sgr A*-IR has ever
been detected at L', by a factor of 3.
From the average images, we find that Sgr A*-IR's positional offsets
from the Galaxy's dynamical center 
are statistically consistent with zero at both K' and L'
([$\Delta_{RA, K'}= -2 \pm 4 ~ mas$, $\Delta_{Dec, K'}= 0 \pm 3 ~ mas$] and
[$\Delta_{RA, L'}= -4 \pm 2 ~ mas$, $\Delta_{Dec, L'} = -7 \pm 3 ~ mas$],
using the Ghez et al. 2005 determination of the dynamical center).
From the individual images, there is no evidence for variation in the 
location of SgrA*-IR 
as a function intensity; previous claim
of such positional variations from Cl\'enet et al. (2005) at L' are 
easily attributable 
to the nearby extended, very red source discussed above.
Since both the photometric and astrometric measurements were made with a 
PSF-fitting routine and since all the known neighboring sources as well as
a background, which is allowed to have low spatial frequency structure,
are part of the model, our reported values should be free of bias from the 
surrounding stellar population and background emission.

\section{DISCUSSION}

\subsection{Extended Red Emission in the Vicinity of Sgr A*}

There are several possible explanations for the extended red source 
located 90 mas South-West of Sgr A*.
First, one might consider that it consists of multiple red stars that 
are blended in the image.  However, we expect the old stellar 
population to be centered on Sgr A*, so if this were the explanation, 
the displacement would have to be due to the chance placement of a 
relatively small number of objects, presumably giants.  However, even 
giants are not as red as the extended red emission, unless they are 
subject to additional extinction beyond that suffered by Sgr A*. 
This 
would then have to correspond to an unlikely placement of multiple 
giants remarkably close in projection to Sgr A*, yet located in the 
background behind additional extinction.  A second possibility is that 
this emission is from an off-center region within an accretion disk 
around Sgr A*.  Off-center brightening in an accretion disk has been 
predicted by Nayakshin et al. (2004) as a response to 
the passage of stars through the accretion disk, but this mechanism 
gives rise to an unresolved source, rather than to the observed, 
extended emission, so one would have to posit multiple recent star-disk 
passages in this particular region, which is also rather unlikely.

The third explanation is that the extended red source is
attributable to thermal emission from a dust cloud.
Indeed, there is a dust feature projected near Sgr A* in 
this direction, as can be seen in mid-IR images (Stolovy et al. 1996).  
It is tempting to hypothesize that this 
feature lies close to Sgr A* and that energy from Sgr A* and its 
immediate entourage of early-type stars heats the dust, giving rise to 
a temperature gradient that declines away from Sgr A*.  
However, if the extended dust feature is close
enough to Sgr A* to be heated by it, it is problematical that it has not
already been sheared by the tidal interaction into a ring around
Sgr A*.  Furthermore, for an intact object, we
would expect a displacement on the order of 60 mas yr$^{-1}$ -- far
larger than our limit of $\sim$ 10 mas yr$^{-1}$ -- if the distance of
the extended red feature from SgrA* were comparable to its projected distance of
94 mas.  It is therefore quite
likely that the extended red dust feature is projected
along the line of sight toward Sgr A*, and that it is heated by an
embedded or a nearby star.

Only a small amount of dust is required to generate the observed emission.
The 3.8-$\mu$m opacity of the dust cloud must
be quite small, given its size and surface brightness:
$\tau(T_d)~=~10^{-6}\times(exp(3787/1200K)~-~1)/(exp(3787/T_d)~-~1)$.
Here, the dust temperature, $T_d$, is referenced to its upper limit of 1200 K,
derived from the dereddened limit on the K'-L' color, and assuming an emissivity
dependence Q$_{\lambda}~\propto~\lambda^{-1}$.
With these assumptions, the bolometric luminosity of the source
can be crudely estimated:
L =  1.2 L$_{\odot}$ (T$_d$/1200)$^5~\tau(T_d)/\tau(1200 K)$.
Assuming further that the source depth is equal to its width on the sky, and
that the gas-to-dust ratio is 100, we find a total source mass equal to
1.3$\times10^{-10}~M_{\odot}~\tau(T_d)/\tau(1200 K)$, clearly a very
modest clump.

\subsection{Infrared Color of Sgr A*}

The radiative emission emerging from the vicinity of 
our Galaxy's supermassive black hole has puzzled
modelers for three decades.  
Its total luminosity
is only 10$^{36}$ ergs s$^{-1}$, or 10$^{-9}$ of the
Eddington luminosity for a $3.7 \times 10^6 M_{\odot}$
black hole, 
most of which is emitted at radio wavelengths.
To account for the low luminosity, current models
rely on radiative inefficiencies in either, or
some combination of, an accretion flow or an outflow
(see, e.g., review by Melia \& Falcke 2001).
Further complicating the picture, 
recent observations have revealed
X-ray and infrared emission, which both show
significant variability at timescales as short
as tens of minutes (Baganoff et al. 2001; Goldwurm et al.
2003; Porquet et al. 2005; Genzel et al. 2003a; Ghez
et al. 2004; Eckart et al. 2004).  While
there is still considerable debate regarding the
physical origin of this variable component, many
of the models have adopted a magnetic mechanism
in which the infrared variability is due to 
synchrotron emission from a temporarily accelerated population
of electrons and the X-ray emission is either 
a continuation of this synchrotron emission or 
synchrotron self-Comptonized emission from
the same electron population that gives rise to
the infrared emission.

\begin{figure}
\epsscale{1.0}
\plotone{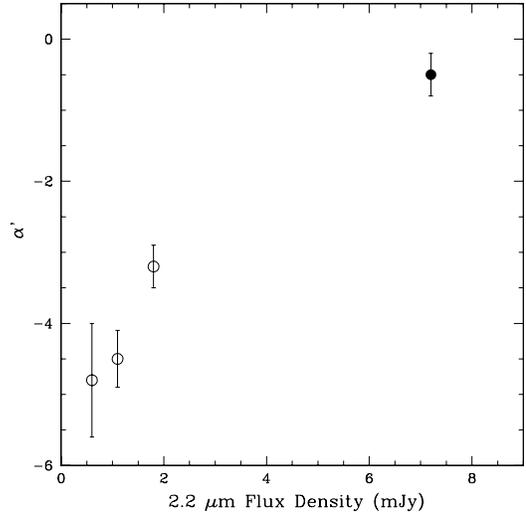}
\figcaption{
Sgr A*'s infrared spectral index ($\alpha$, where $F_{\nu} \sim \nu^{\alpha}$)
as a function of its 2 $\mu$m 
flux density.  Our measurement is the filled point and measurements
from Eisenhauer et al. (2005) are the unfilled points.  We
suggest that the spectral index may depend on the strength of Sgr A*'s
infrared brightness.
\label{alpha}
}
\end{figure}

The results from this paper support the synchrotron
flare models in a number of ways.
First, our observations show that, unlike at 
X-ray wavelengths, the emission at infrared
wavelengths appears to have no true steady-state
component\footnote{It should be noted that the X-ray
steady-state component is spatially resolved with a size
that corresponds to the Bondi radius ($\sim$1$\arcsec$)
and thus comes from scales much larger than are being
probed at infrared wavelengths.}.  
Our L'= 14.7 mag, or de-reddened flux
density of 1.3 mJy, detection of Sgr A*-IR is a factor
of 3 fainter than any previous detection.  So while
temporarily stable flux densities (the so-called 
``interim quiescent" values) have been reported by
Genzel et al. (2003) and Eckart et al. (2004), there appears to be 
no long-term true steady state at infrared wavelengths.
Here it is worth adding a cautionary note regarding
the results that are derived from aperture photometry;
in such a crowded field apparent steady-states can easily
be created from stellar confusion, the level of which will change
on roughly month timescales due to the motions of stars
as they undergo periapse passage.  If the flares are
caused by magnetic field reconnection or some other local
heating event that generates a highly accelerated electron 
population, there is no predicted steady-state component.

A second source of support for synchrotron flare models 
from this work is the observed infrared colors, which
are predicted and observed to be red.  However it
is interesting to note that the infrared color
reported here and by Eisenhauer et al. (2005), 
while both red, are significantly different from each other.  
Our observed K'-L' color of 3.0 $\pm$ 0.2 corresponds
to a de-reddened value of 1.4 $\pm$ 0.2, which if modeled as a power
law ($F_{\nu} = \nu^{\alpha}$) gives $\alpha$ = -0.5 $\pm$ 0.3.
As shown in Figure ~\ref{alpha}, this is substantially bluer than the 
spectral index
obtained by Eisenhauer, $\alpha \sim -4 \pm 1$, from spectral
observations between 2.0 and 2.4 $\mu$m.  While these spectral
indices are inferred from different types of measurements, it is
difficult to invoke a calibration error, such as improper re-reddening or
data extraction, or the 1 minute time lag in our observations, that is
large enough to account for the discrepant spectral indices.
It is however straight forward to invoke plausible physical differences.  
A spectral break between 2.4 and 3.8 $\mu$m could account
for the different inferred spectral indices.  Such a spectral break 
is necessary to invoke
between the Eisenhauer measurement and the de-reddened 
mid-infrared limits from Sgr A*
reported by Morris (2001; 50 mJy at 12.5 $\mu$m, 70 mJy at 20.8 $\mu$m, and
85 mJy at 24.5 $\mu$m), since a constant power-law extrapolated
from our lowest L' measurement would imply extremely
large mid-infrared fluxes (150 - 2300 mJy).  However, this is not
necessary for our values, 
which predict mid-infrared fluxes of only $\sim$ 2 - 3 mJy, well below
the current mid-infrared limits, from
extrapolations of our lowest L' measurement.   
Furthermore, while such large spectral breaks might be reasonable
to invoke between near-infrared and mid-infrared wavelengths,
they are rather difficult to invoke over the small wavelength
difference between our experiment and that of Eisenhauer et al. (2005).
More intriguingly, as an explanation, the spectral index
derived by this work was obtained from measurements taken
when Sgr A*-IR was a factor of $\sim$6 brighter than it was
during the Eisenhauer measurements (see Figure \ref{alpha}).  
Thus a possible cause for the discrepancy may be an 
infrared spectral index that depends on the strength of Sgr A*-IR's emission.
Such a dependence could suggest that events that generate
a higher fraction of energetic electrons are responsible
the stronger infrared flares. 

\section{SUMMARY}

We have obtained the first observations of the Galactic center with
a Laser Guide Star Adaptive Optics system,
demonstrating the power of this technique.  It
has produced the sharpest L'(3.8 $\mu$m) images yet of this region.
Furthermore, compared to Keck's NGS-AO performance on the
Galactic center, LGS-AO results in L' Strehl ratios that are a factor of
two higher and a factor of four to five more stable.  This
imaging performance enhancement has permitted us to clearly image
an extended infrared emission feature located a mere 90 mas from
the central black hole.  This object is likely to be a relatively hot
dust feature that is locally heated by an embedded or nearby star
and that is projected along the line of sight toward Sgr A*.
In addition, with improved
image quality, Sgr A* can be reliably detected in much shorter
time exposures, which has permitted a measurement of Sgr A*'s infrared
color.  In our measurement, Sgr A* is much bluer and brighter than was
measured by Eisenhauer et al. (2005).  Among the many possible  
interpretations
for this discrepancy, one intriguing possibility is that Sgr A*'s  
infrared
spectral index is variable and that events leading to
stronger infrared flares generate a higher fraction of energetic
electrons.  In conclusion, LGS-AO has improved the quality
and versatility with which diffraction-limited observations
can be made at the Keck telescope and is ushering in an exciting
new era of observations of the Galactic center.

\acknowledgements

The authors thank observing assistants Cynthia Wilburn 
and Gary Puniwai for their help in obtaining the observations
as well as Reinhard Genzel and Quinn Konopacky for useful comments on the 
manuscript.
Support for this work was provided by NSF
grant AST-0406816 and the NSF Science
\& Technology Center for AO, managed by UCSC
(AST-9876783), and the Packard Foundation.
The W. M. Keck Observatory, 
is operated as a scientific partnership among the California Institute 
of Technology, the University of California and the National Aeronautics and 
Space Administration.  The Observatory was made possible by the generous 
financial 
support of the W. M. Keck Foundation.  The authors wish to recognize and 
acknowledge the very significant cultural role and reverence that the summit 
of Mauna Kea has always had within the indigenous Hawaiian community.  We are 
most fortunate to have the opportunity to conduct observations from this 
mountain.

\pagebreak

\end{document}